\documentclass[iop,apjl]{emulateapj}
\usepackage{color} 

\usepackage{ulem}
\usepackage{amsmath,apjfonts}
\usepackage{enumerate}

\newcommand\cts{counts~s$^{-1}$}
\newcommand\ergs{erg~s$^{-1}$}
\newcommand\ergcms{erg~cm$^{-2}$~s$^{-1}$}
\newcommand\flam{erg~cm$^{-2}$~s$^{-1}$~\AA$^{-1}$}

\newcommand{\halpha}{H{$\alpha$}}

\newcommand{\HI}{H\,{\sc i}}
\newcommand{\HII}{H\,{\sc ii}}

\begin{document}

\title{A Luminous X-ray Flare From The Nucleus of The Dormant Bulgeless Spiral Galaxy NGC 247}

 \author{
 Hua Feng\altaffilmark{1,2}, 
 Luis C.~Ho\altaffilmark{3,4}, 
 Philip Kaaret\altaffilmark{5}, 
 Lian Tao\altaffilmark{6,7,2}, 
 Kazutaka Yamaoka\altaffilmark{8},
 Shuo Zhang\altaffilmark{9},
 Fabien Gris\'{e}\altaffilmark{10}}

\altaffiltext{1}{Department of Engineering Physics, Tsinghua University, Beijing 100084, China}
\altaffiltext{2}{Center for Astrophysics, Tsinghua University, Beijing 100084, China}
\altaffiltext{3}{Kavli Institute for Astronomy and Astrophysics, Peking University, Beijing 100871, China}
\altaffiltext{4}{Department of Astronomy, School of Physics, Peking University, Beijing 100871, China}
\altaffiltext{5}{Department of Physics and Astronomy, University of Iowa, Iowa City, IA 52242, USA}
\altaffiltext{6}{Cahill Center for Astronomy and Astrophysics, California Institute of Technology, Pasadena, CA 91125, USA}
\altaffiltext{7}{Department of Physics, Tsinghua University, Beijing 100084, China}
\altaffiltext{8}{Department of Physics and Mathematics, Aoyama Gakuin University, 5-10-1 Fuchinobe, Chuo-ku, Sagamihara 252-5258}
\altaffiltext{9}{Columbia Astrophysics Laboratory, Columbia University, New York, NY 10027, USA}
\altaffiltext{10}{Observatoire astronomique de Strasbourg, Universit\'{e} de Strasbourg, CNRS, UMR 7550, 11 rue de l'Universit\'{e}, F-67000 Strasbourg, France}

\shorttitle{A Luminous X-ray Flare from the NGC 247 Nucleus}
\shortauthors{Feng et al.}

\begin{abstract}
NGC 247 is a nearby late-type bulgeless spiral galaxy that contains an inactive nucleus. We report a serendipitous discovery of an X-ray flare from the galaxy center with a luminosity up to $2 \times 10^{39}$ \ergs\ in the 0.3--10 keV band with {\it XMM-Newton}. A {\it Chandra} observation confirms that the new X-ray source is spatially coincident with the galaxy nucleus. The {\it XMM-Newton} data revealed a hard power-law spectrum with a spectral break near 3--4 keV, no pulsations on timescales longer than 150 ms, and a flat power spectrum consistent with Poisson noise from 1 mHz to nearly 10 Hz. Follow-up observations with {\it Swift} detected a second flux peak followed by a luminosity drop by factor of almost 20. The spectral and temporal behaviors of the nuclear source are well consistent with the scenario that the flare was due to an outburst of a low-mass X-ray binary that contains a stellar-mass black hole emitting near its Eddington limit at the peak. However, it cannot be ruled out that the sudden brightening in the nucleus was due to accretion onto a possible low-mass nuclear black hole, fed by a tidally disrupted star or a gas cloud;  the {\it MAXI} observations limit the peak luminosity of the flare to less than $\sim10^{43}$~\ergs, suggesting that it is either a low mass black hole or an inefficient tidal disruption event (TDE).
\end{abstract}

\keywords{galaxies: nuclei --- X-rays: galaxies --- black hole physics}

\section{Introduction}
\label{sec:intro}

It is widely accepted that most massive galaxies harbor supermassive black holes  \citep[for reviews see][]{Ho2008,Kormendy2013}. However, It is still unknown what fraction of small, bulgeless galaxies contain nuclear black holes \citep[for a review see][]{Greene2012}. If they do, they are most likely to harbor low-mass black holes with masses $\lesssim 10^6$~$M_\sun$ or even less \citep[e.g.,][]{Filippenko2003,Barth2004,Peterson2005,Seth2010,Reines2011}. Current searches for nuclear black holes in small galaxies were mainly done in optical \citep{Greene2004,Greene2007b,Barth2008,Izotov2008,Izotov2012,Dong2012}. These means are of low efficiency because they only allow us to find the most active nuclei in nearby galaxies.  X-ray observations help to detect X-ray cores with much lower levels of activity, but the physical nature of the X-ray cores is uncertain \citep{Desroches2009,Lemons2015}. In the meanwhile, the active fraction goes down toward the low-mass end for nuclear black holes \citep{Greene2007a}. 

For the majority of inactive low-mass galaxies, the tidal disruption event \citep[TDE;][]{Rees1988} could be a powerful probe for their discovery and in-depth study.  TDEs may allow us to census black holes in dormant galaxies and even estimate the black hole mass and spin with detailed X-ray spectroscopy and timing. A couple of dozens of such events or suspected events have been reported to date \citep[for a review see][]{Komossa2012}. The TDE rate was estimated to be of the order of $10^{-5}$ to $10^{-4}$ per year per galaxy, based on possible detections in X-ray surveys with {\it ROSAT} \citep{Donley2002}, {\it XMM-Newton} \citep{Esquej2007}, or {\it Chandra} \citep{Luo2008}. Thus it would have been very lucky to catch a TDE in or around our Local Group. The chance may be enhanced in low-mass galaxies, as theoretically the TDE rate is inversely proportional to the black hole mass, and becomes the highest in nucleated dwarf galaxies if they contain black holes \citep{Wang2004}. Indeed, two candidate TDEs have been detected in dwarf galaxies, WINGS~J1348 in the galaxy cluster Abell 1795 \citep[$z = 0.06$;][]{Maksym2013,Maksym2014b,Donato2014} and RBS~1032 associated with SDSS J114726.69+494257.8 \citep[$z = 0.026$;][]{Maksym2014a}, both located more than 100~Mpc away.

Here, we report a serendipitous detection of a strong X-ray flare from the nucleus of NGC 247, which is a nearby dormant galaxy. If it is indeed a nuclear black hole event, then this source offers us a unique case besides our Galactic center for studying accretion onto quiescent black holes due to its proximity, and if it is a TDE, will allow us to investigate the TDE behavior in its late phase. On the other hand, the flare could be due to an outburst from an unresolved stellar-mass black hole in the nuclear star cluster. Such objects were excluded in the study of black hole binaries and/or ultraluminous X-ray sources (ULXs) in external galaxies, almost without exception. In this regard, some of these objects, such as the X-ray cores of M33 \citep{LaParola2003} and NGC 253 \citep{Brunthaler2009}, could be misidentified as weak X-ray active galactic nuclei (AGNs) especially with the lack of multiwavelength information, though their contamination to AGN surveys is estimated to be low \citep{Desroches2009,Gallo2010}. 

NGC 247 is a nearby late-type spiral galaxy classified as SAB(s)d \citep{deVaucouleurs1991} in the Sculptor group, which is just outside of our Local Group.  We adopt an IR-based Cepheid distance of 3.4 Mpc to the galaxy \citep{Gieren2009}. The total \HI\ mass was measured to be $\sim$$10^9$~$M_\sun$ \citep{Carignan1990,Ott2012}. The galaxy has a high inclination angle, 74~degrees, and an isophotal diameter $D_{25} = 19.7$~kpc assuming the distance mentioned above \citep{Carignan1985}. It has a total apparent $B$ magnitude of about 9.7 \citep{Carignan1985,Pierce1992}.  Assuming a Galactic foreground extinction $A_B = 0.078$ \citep{Schlegel1998}, an internal extinction $A_B = 0.67$ \citep{Carignan1985}, and a distance modulus of 27.64 \citep{Gieren2009}, the total absolute blue magnitude is $M_B^0 = -18.7$.  

NGC 247 contains a nuclear star cluster similar to other late-type spiral galaxies \citep{Davidge2002}. The dynamical center of the galaxy determined via 21 cm observations was found to be consistent with its optical center \citep{Carignan1990}. UV spectroscopy of the nuclear cluster \citep{Maoz1995} suggests that it is an \HII\ region with no signature of an active galactic nucleus (AGN).  Past X-ray observations also indicate that it is a dormant galaxy. 

The star formation rate of the galaxy is estimated to be 0.03 $M_\sun$~yr$^{-1}$ using the \halpha\ emission line \citep{Ferguson1996}.  The low star formation rate, and the lack of tidal features in NGC 247 and \HI\ streams between NGC 247 and its neighbor NGC 253 \citep{Davidge2010}, suggest that there is no ongoing or past mergers that could potentially trigger the activity of an AGN.

\section{Observations}
\label{sec:obs}

\subsection{Previous non-detections}

Before 2014, X-ray observations of NGC 247 did not detect any X-ray emission from the nuclear region. A {\it Chandra} observation was made on 2011 February 01 (ObsID 12437) with an effective exposure of 5.0 ks. We created a new level 2 event file using the {\tt repro} script in CIAO. We calculated a radius of 1.573\arcsec\ that enclosed 90\% of the PSF in the nuclear position using the {\tt psfsize\_srcs} command and adopted it as the source aperture. A concentric annular region was used to estimate the background.  There was actually no photon detected in the source aperture in 0.3--8 keV. The 90\% upper limit of the source flux, based on background fluctuation in the source aperture assuming Poisson distribution, was estimated to be $5.10 \times 10^{-4}$ \cts\ in 0.3--8 keV using the Bayesian approach with the {\tt aprates} command. With the local spectral response files, assuming a broken power-law spectrum that was obtained during the flare (see the next paragraph and Table~\ref{tab:fit}), such a count rate can be translated to a 90\% flux upper limit of $1.1 \times 10^{-14}$ \ergcms\ or an absorption-corrected luminosity of $1.8 \times 10^{37}$ \ergs\ in 0.3--10 keV. 

\begin{figure}[t]
\centering
\includegraphics[width=\columnwidth]{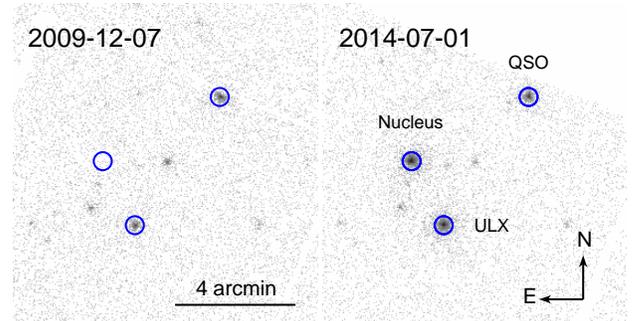}
\caption{{\it XMM-Newton} MOS1 images of the galaxy NGC 247 in 2009 versus 2014.  The galaxy used to show a quiescent nucleus but a new, bright X-ray source was detected in 2014 and confirmed to be coincident with the galaxy nucleus with a follow-up {\it Chandra} observation. The ULX and QSO were used to align {\it Chandra} and {\it HST} images. 
\label{fig:xmm}}
\end{figure}

The galaxy was observed with {\it XMM-Newton} on 2009 December 07 (ObsID 0601010101, see Figure~\ref{fig:xmm}). We created new event files with updated calibration files, and selected events from intervals where the background flux was within $\pm 3 \sigma$ of the mean quiescent level. The PN data were not used because the nuclear position was located on bad CCD columns. The MOS1 and MOS2 data, with an effective exposure of nearly 30 ks, were used to estimate the detection limit with a source and a larger, nearby background region.  A source aperture size of 20\arcsec\ in radius was found to be the optimal choice. As there are sufficient counts in the aperture, it allows us to use the normal distribution to approximate the Poisson distribution. With spectral responses generated in the nuclear position and assuming the same broken power-law model (see Table~\ref{tab:fit}), we estimate a 90\% upper limit of the observed flux as $4.9 \times 10^{-15}$ \ergcms\ and an absorption-corrected luminosity of $8.0 \times 10^{36}$ \ergs\ in 0.3--10 keV.  Another {\it XMM-Newton} observation made on 2001 July 08 does not enable us to place  a stringent upper limit due to heavy background contamination.

\subsection{First detection with {\it XMM-Newton}}

\begin{deluxetable*}{llllllccccl}
\tablecolumns{11}
\tabletypesize{\scriptsize}
\tablewidth{0pc}
\tablecaption{X-ray spectral fitting to the {\it XMM-Newton} spectrum.
\label{tab:fit}}
\tablehead{
\colhead{Model} & \colhead{$N_{\rm H,ext}^{a}$} & \colhead{$\Gamma$} & \colhead{$N_{\rm  PL}^b$}  &  \colhead{$E_{\rm b}^c$} & \colhead{$\Gamma_2$/$p^d$} &  \colhead{$T_{\rm in}$}  & \colhead{$R_{\rm in}^e$} & \colhead{$f_{\rm 0.3-10\;keV}$} & \colhead{$L_{\rm 0.3-10\;keV}$} & \colhead{$\chi^2/{\rm dof}$} \\ 
\colhead{} & \colhead{($10^{21}$~cm$^{-2}$)} & \colhead{} & \colhead{} & \colhead{(keV)} & \colhead{} & \colhead{(keV)} & \colhead{(km)} & \colhead{($10^{-12}$ \ergcms)} & \colhead{($10^{39}$ \ergs)} & \colhead{}
}
\startdata
power-law & $1.50 \pm 0.12$ & $1.74 \pm 0.04$ & $2.39 \pm 0.11$ & \nodata & \nodata & \nodata & \nodata & $1.29 \pm 0.03$ & $2.22 \pm 0.05$ & 549.04/490 \\

broken PL & $1.09 \pm 0.15$ & $1.50 \pm 0.08$ & $2.07 \pm 0.12$ & $3.5 \pm 0.4$ & $2.3 \pm 0.2$ & \nodata & \nodata & $1.23_{-0.02}^{+0.05}$& $1.98 \pm 0.07$ & 490.79/488 \\

cutoff PL & $0.78 \pm 0.18$ & $1.07 \pm 0.15$ & $2.35 \pm 0.10$ & $4.7_{-0.9}^{+1.4}$ & \nodata & \nodata & \nodata & $1.22_{-0.02}^{+0.03}$ & $1.87 \pm 0.08$ & 494.63/489 \\

MCD & $0.18 \pm 0.06$ & \nodata & \nodata & \nodata & \nodata & $1.55 \pm 0.05$ & $34 \pm 2$ & $1.15_{-0.03}^{+0.04}$ & $1.63 \pm 0.04$ & 573.00/490 \\

$p$-free & $0.95 \pm 0.17$ & \nodata & \nodata & \nodata & $0.59 \pm 0.02$ & $2.3 \pm 0.3$ & $11 \pm 3$ & $1.21_{-0.02}^{+0.04}$ & $1.91 \pm 0.08$ & 491.89/489 \\

PL + MCD & $1.5_{-0.5}^{+0.8}$ & $2.3_{-0.5}^{+0.8}$ & $1.4_{-0.3}^{+0.4}$ & \nodata & \nodata & $1.92_{-0.33}^{+0.16}$ & $19_{-3}^{+4}$ & $1.21_{-0.02}^{+0.03}$ & $2.1_{-0.2}^{+0.6}$ & 490.18/488 
\enddata
\tablecomments{All errors are quoted in 90\% confidence level.}
\tablenotetext{a}{Extragalactic absorption column density; an additional absorption component fixed at $N_{\rm H} = 2.06 \times 10^{20}$~cm$^{-2}$ to account for Galactic absorption is not shown here.}
\tablenotetext{b}{Normalization in units of $10^{-4}$~photons~keV$^{-1}$~cm$^{-2}$~s$^{-1}$ at 1~keV.}
\tablenotetext{c}{Break energy for the broken power-law model (bknpower in XSPEC) or the e-folding energy for the cutoff power-law model (cutoffpl in XSPEC).}
\tablenotetext{d}{$\Gamma_2$ is the power-law photon index above the break energy for the broken power-law model; $p$ is the radial dependence of the disk temperature in the $p$-free model.}
\tablenotetext{e}{Assuming a face-on disk.}

\end{deluxetable*}

The first detection of a bright X-ray source in the nuclear region of NGC 247 was made on 2014 July 01 with {\it XMM-Newton} (ObsID 0728190101), see Figure~\ref{fig:xmm}. Following the same process mentioned above, we created new event files and filtered for background flares, resulting in an effective exposure of 24.3~ks. The X-ray energy spectra were extracted from a circular region with a radius of 32 arcsec, and the background spectra were extracted from a nearby circular region on the same CCD chip, using events with FLAG $=$ 0 and PATTERN $\le$ 12 for MOS and $\le 4$ for PN. The spectra were grouped such that each new spectral bin was 1/4 of the local FWHM or had at least 15 net counts, from 0.2 to 10 keV for PN and from 0.3 to 10 keV for MOS.  The PN and MOS spectra were loaded into XSPEC v12 jointly. A multiplicative constant for MOS flux relative to the PN flux was introduced as a free parameter to account for possible systematics in their effective area calibration, and was found to be close to 1.05. The interstellar absorption consists of a Galactic foreground component fixed at $2.06 \times 10^{20}$~cm$^{-2}$ \citep{Kalberla2005} and a free, extragalactic component, both using the TBabs model \citep{Wilms2000}. 

\begin{figure}[thb]
\centering
\includegraphics[width=\columnwidth]{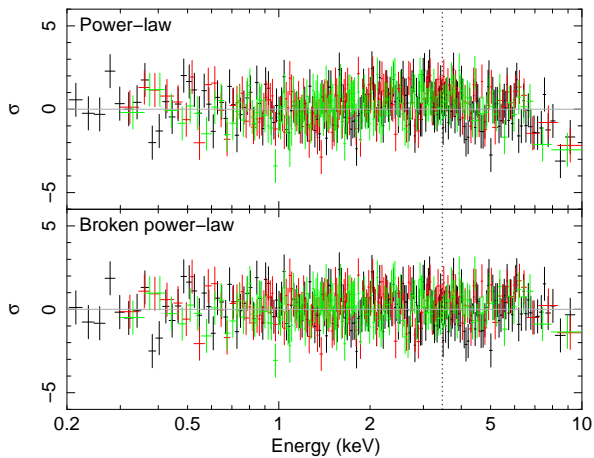}
\caption{Comparison of the X-ray spectral fitting with the power-law model ($\chi^2 = 549.0$ with 490 degrees of freedom) and the broken power-law model ($\chi^2 = 490.8$ with 488 degrees of freedom). The dotted line indicates the energy of the spectral break. The residuals after fitting with the power-law model show a curved feature in the 2--10 keV range.
\label{fig:fit}}
\end{figure}

We first tried a simple power-law model to fit the spectrum. The fitting is inadequate (a null hypothesis probability of 0.03 while we define $>0.05$ being acceptable) and a trend of spectral break can be seen in the residual above a few keV.  We thus tried a broken power-law model to account for the possible break, which significantly improved the fits and the presence of the break has a significance of $1.3 \times 10^{-12}$ with partial F-test. A comparison of the power-law and broken power-law fits is displayed in Figure~\ref{fig:fit}.  An exponentially cut-off power-law model produces a similar goodness of fit. We then tried physical models that allowed for a high-energy cutoff, including the multicolor disk (MCD) model as an approximation to the standard accretion disk model \citep{Shakura1973} and the $p$-free model, with the radial dependence of temperature ($T(R) \propto R^{-p}$) being a free parameter, as an approximation to the slim-disk model \citep{Mineshige1994}. We also tested a double-component model, MCD plus power-law, which is the standard to fit the X-ray energy spectra of accreting black hole binaries.  The spectral parameters along with observed fluxes and absorption-corrected luminosities (computed with the cflux model) are listed in Table~\ref{tab:fit}.  The power-law and MCD models cannot provide an adequate fit (null hypothesis probability < 0.05), while the other models can. 

\begin{figure}[t]
\centering
\includegraphics[width=0.8\columnwidth]{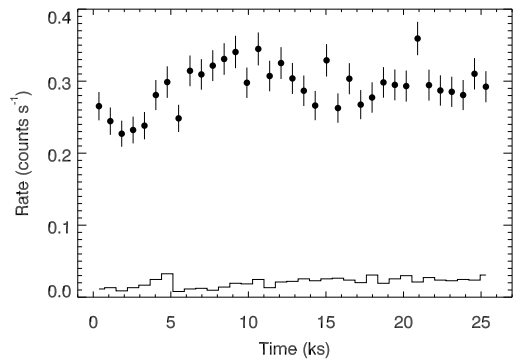}
\caption{X-ray light curve of the nuclear source. Points are the net count rate and the histogram is the estimated background contribution in the source aperture. The time step is 733.6 s, a binning factor of $10^4$ on the CCD frame time. The variation is reflected on the periodogram or the power spectrum as a single peak at the timescale of 13.5 ks. A constant fit results in $\chi^2 = 82.2$ with 34 degrees of freedom, corresponding to a chance probability of $7.2 \times 10^{-6}$.
\label{fig:xmmlc}}
\end{figure}

The PN data have a continuous good time interval of 26.1 ks (no background flares or timing gaps), allowing us to search for timing signals.  Coherent signals were searched in the full band (0.2--10 keV) based on the CCD frame time (about 73 ms) with a 35514-point Fourier transform. The highest power amplitude is 31.5 at 0.076680 mHz under Leahy normalization \citep{Leahy1983}, while the 99\% confidence level corresponds to an amplitude of 34.8 assuming $\chi^2$ distribution with 2 degrees of freedom and taking into account the number of trials. With the $Z_1^2$-test \citep{Buccheri1983}, the detected amplitude allows us to place an upper limit on the pulsed fraction to be 4.3\% assuming a sinusoidal signal. Thus, no pulsations on timescales larger than $\sim$150 ms are found in the data. Searched in different timescales and energy ranges, we did not detect quasi-periodic oscillations or any form of excessive timing noise above the Poisson level. The power spectra in the frequency range from $10^{-3}$--10 Hz are well consistent with Poisson noise. The only possible temporal variation is seen on a timescale near 13.5 ks (Figure~\ref{fig:xmmlc}). However, limited by the total duration of the lightcurve, it is hard to interpret the possible variation. 

\subsection{Fine position with {\it Chandra}}

On 2014 November 12, when the source was still active, we conducted a {\it Chandra} observation in order to precisely locate its position, to see whether the bright X-ray source found by {\it XMM-Newton} is coincident with the nucleus of the galaxy.  The observation was done with the ACIS-S detector and designed to include the new source, an ultraluminous X-ray source (ULX; 2.4\arcmin\ to its southwest), and a background QSO PHL 6625 (4.5\arcmin\ to its northwest) \citep{Jin2011} on the same CCD chip. Both the ULX and the QSO were identified to have a unique optical counterpart \citep{Tao2012}, enabling us to align the {\it Chandra} and {\it HST} images and improve the relative astrometry.  We created exposure-map corrected images and performed source detection with the {\tt wavdetect} command.  The source spectrum was extracted using the {\tt specextract} script, grouped to be at least 15 counts per energy bin, and fitted with an absorbed power-law model (in the same format as for {\it XMM-Newton}) in the energy range of 0.3--8.0 keV. The simple power-law model gives an adequate fit ($\chi^2 = 26.57$ with 25 degrees of freedom), with an extragalactic absorption column density $N_{\rm H} = (3.6 \pm 1.8) \times 10^{21}$~cm$^{-2}$, a photon index $\Gamma = 1.9 \pm 0.3$, an observed flux of $(1.14 \pm 0.17) \times 10^{-12}$~\ergcms, and an intrinsic luminosity of $2.5_{-0.4}^{+0.6} \times 10^{39}$~\ergs\ in 0.3--10 keV. If we adopt the broken power-law model (Table~\ref{tab:fit}) and leave only the normalization as a free parameter,  we derive $f_{\rm X} = (1.10 \pm 0.09) \times 10^{-12}$~\ergcms\, and $L_{\rm X} = (1.81 \pm 0.14) \times 10^{39}$~\ergs\ in 0.3--10 keV, with $\chi^2 = 30.35$ and 27 degrees of freedom. All the errors are at the 90\% confidence level. 

\begin{figure}[t]
\centering
\includegraphics[width=\columnwidth]{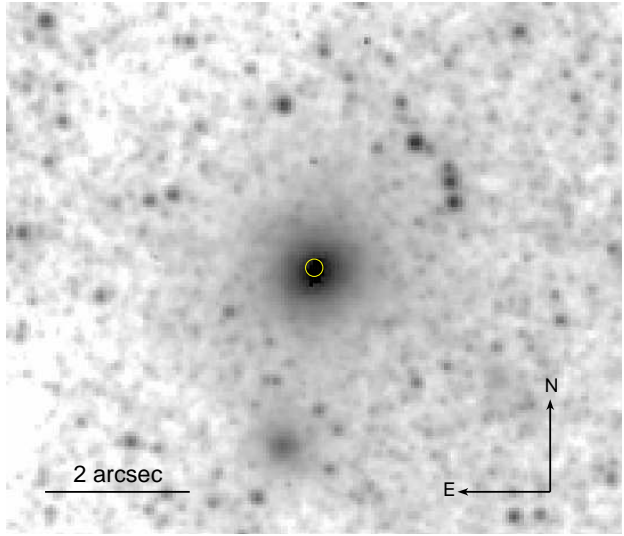}
\caption{HST ACS WFC F606W image around the NGC 247 nucleus with the corrected X-ray position of the nuclear source, which has an error radius of 0.12\arcsec.  The X-ray source is spatially coincident with the nuclear star cluster.
\label{fig:hst}}
\end{figure}

An HST image covering the three X-ray sources was created using archival ACS WFC F606W images with the {\tt multidrizzle} task.  Using the {\tt geomap} task in IRAF, we registered the X-ray positions of the ULX and the QSO to the {\it HST} positions of their optical counterparts, with shifts in X and Y and a rotation. Then the {\tt geoxytran} task was used to transform the X-ray position of nuclear source onto the HST coordinates. The resultant position uncertainty includes the statistical errors of the X-ray positions from {\tt wavdetect} and the coordinate transformation errors from the {\tt geomap} fitting, and was calculated to be 0.12\arcsec\ in 90\% confidence level (assuming a Rayleigh distribution). We note that very similar result is obtained if the alignment is done only with the ULX, which is closer to the nucleus. The corrected position of the nuclear source on the {\it HST} image is displayed in Figure~\ref{fig:hst}.

\subsection{Follow-up monitoring with {\it Swift} and the long-term lightcurve}

We started a weekly monitoring program for the activity of the nuclear source with the {\it Swift} X-ray telescope (XRT) from 2014 October 13 to 2015 May 19. The source was unobservable with Swift for three months towards the end of the monitoring due to Sun avoidance.  The effective exposure of each observation varied from 1.3 to 2.5 ks.  For each observation, we extracted the source spectrum from an aperture of 20 pixel radius, and a background spectrum from a nearby source-free region, using the event file delivered from the pipeline.  The detected number of photons in the source aperture ranged from 4 to 52, and the background contribution was less than 40\% in the faintest case. Each spectrum was grouped into two energy bins, one in 0.3--2 keV and the other in 2--10 keV. To estimate the source luminosity, we fitted every spectrum in the Cash statistics with the broken power-law model derived from {\it XMM-Newton} (Table~\ref{tab:fit}) by fixing all parameters except the normalization. The absorption-corrected luminosity was computed with cflux in 0.3--10 keV.  If the source was not detected above 1$\sigma$, its 90\% upper limit was calculated using the same Bayesian approach described above.

\begin{figure}[tb]
\centering
\includegraphics[width=\columnwidth]{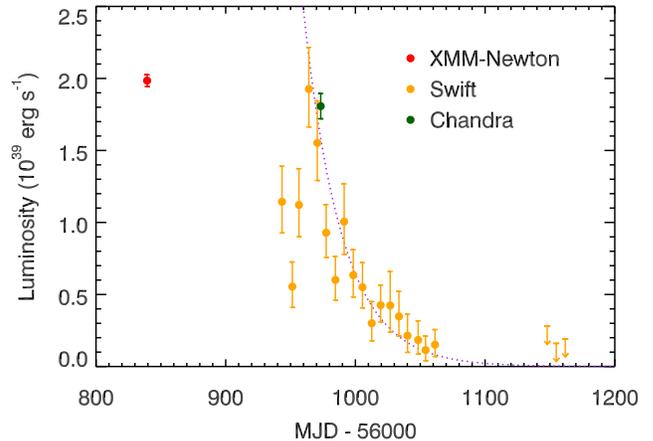}
\caption{0.3--10 keV X-ray lightcurve of the NGC 247 nucleus. The dashed line is an exponential decay function, $L = L_0\exp [-(t - t_0) / \tau]$, fitted to the data after MJD 56960. 
\label{fig:lc}}
\end{figure}

The X-ray lightcurve (Figure~\ref{fig:lc}) displayed a luminosity change by a factor of 17. To be consistent, we adopted the {\it Chandra} luminosity that was computed assuming the broken power-law model.  The irregular variation of the source luminosity and the short baseline do not allow us to fit a power-law decay model to the lightcurve. The {\it Swift} lightcurve appears like a fast rise and exponential decay,  typical of the outburst behavior of an accreting low-mass X-ray binary. We therefore fitted an exponential decay function, $L = L_0\exp [-(t - t_0) / \tau]$, to the data after the MJD 56960 and derived a decay timescale of ($30 \pm 3$) d. 

Along with the XRT monitoring, the UVOT was in operation with filters UVW1, UVW2, and UVM2 by turns. The source displayed a constant UV flux. The mean flux density corrected for Galactic reddening \citep[$E(B-V) = 0.018$;][]{Schlegel1998} is $1.3 \times 10^{-15}$ \flam\ in UVW2 (central wavelength at 1928\AA), $1.2 \times 10^{-15}$ \flam\ in UVM2 (at 2246\AA), and $9.7 \times 10^{-16}$ \flam\ in UVW1 (at 2600\AA), measured via a source aperture of 3\arcsec\ and multiple circular background apertures surrounding it.  The total luminosity in these UV bands ($\sim$1500-3000\AA) is roughly $3 \times 10^{39}$ \ergs.  These UV fluxes are consistent with those measured with the Optical Monitor onboard {\it XMM-Newton} and arise from the nuclear star cluster \citep{Maoz1995}, suggesting that no detectable UV emission was found to be associated with the X-ray activity. 

\subsection{Search in the {\it MAXI} archive}

We searched X-ray emission from the NGC 247 direction with the {\it Monitor of All-sky X-ray Image} ({\it MAXI}) onboard the international space station \citep{Matsuoka2009}. The search was done with the gas slit camera (GSC) and the solid-state slit camera (SSC) in a variety of energy bands (e.g., 0.7--4, 2--4, 2--6, or 2--10 keV) and timescales (e.g., 1 day, 7 days, or 30 days), in the time window before the first detection with {\it XMM-Newton} and the previous non-detection with {\it Chandra} (MJD 55594--56839). However, no detection over a significance of 3$\sigma$ can be found with {\it MAXI}. The non-detection allows us to place a 3$\sigma$ flux upper limit on the peak flux averaged over one week of $5 \times 10^{-11}$ \ergcms\ in 2--10 keV (with GSC) or $3 \times 10^{-10}$ \ergcms\ in 0.7--2 keV (with SSC). Provided the broken power-law spectrum measured with {\it XMM-Newton}, the upper limit in flux can be translated to an upper limit in luminosity in 0.3--10 keV of a few times $10^{41}$ \ergs\ at a distance of NGC 247 corrected for absorption. If we assume a softer spectrum, e.g., a disk blackbody with an inner temperature of 0.1 keV and an absorption column density of $10^{21}$~cm$^{-2}$, the upper limit in luminosity could be as high as $3 \times 10^{43}$~\ergs\ in the energy band of 0.3--10 keV, constrained by the SSC upper limit. 

\section{Discussion}
\label{sec:dis}

In this paper, we report the detection of a luminous X-ray flare from a compact object in the nuclear star cluster (also the dynamical center) of the nearby late-type galaxy NGC 247, which is known to be inactive in both optical and X-ray. There are two possible interpretations of such a sudden brightening in a galaxy nucleus: due to an outburst of a low-mass X-ray binary that resides in the nuclear star cluster, or accretion activity onto a nuclear black hole, such as a TDE. 


If not spatially coincident with the galaxy nucleus, this object can be naturally explained as a stellar-mass black hole accreting from a Roche-lobe overflowed low-mass companion star.  The luminosity seems to peak around $2 \times 10^{39}$ \ergs, which is the Eddington limit of a 10--20 solar mass black hole. A few neutron star binaries have produced an apparent luminosity this high \citep{Skinner1982,Bachetti2014}.  The non-detection of pulsed emission in the XMM-Newton data suggests that it could be a neutron star only if the pulsed fraction is lower than 4\% or the period is shorter than 146 ms. The lightcurve,  especially at times after the peak around MJD 56970, seems to follow a simple exponential decay on a timescale of a month or so, which is often seen in low-mass X-ray binaries as the first phase of flux decay during an outburst \citep{Lasota2001}.  The measured decay timescale is roughly consistent with what we have observed for Galactic black hole binaries \citep[see Figures~5--8 in][]{Remillard2006}.  

The presence of a spectral break at a few keV has been ubiquitously found in bright ULXs and argued as a signature of super-Eddington accretion \citep{Stobbart2006,Gladstone2009,Motch2014}.  Specifically, depending on the MCD plus power-law fitting, this source is typical of a sub-class of ULXs, dubbed as the {\it broadened disk} class by \citet{Sutton2013}, which have a relatively low luminosity (near the luminosity threshold for ULXs) and were interpreted as due to normal stellar-mass black holes ($\lesssim 20$~$M_\sun$) accreting near the Eddington level. We argue that the MCD plus power-law fit is not physical because the power-law component, which is presumably produced by up-scattering of disk photons in an optically-thin corona,  dominates over the MCD component at low energies (below 1.5 keV).  The $p$-free model gives a self-consistent interpretation of its nature, being a slim-disk where advection leads to a flattened radial temperature profile ($p = 0.6$ versus $p = 0.75$ for a standard disk).  The derived temperature and luminosity are consistent with emission from a stellar-mass black hole with near Eddington accretion. The absence of timing noise is also consistent with the spectrum being dominated by emission from a thermal slim-disk. 

According to the Chandra survey of ULXs in nearby galaxies \citep{Swartz2011}, we estimated a probability of roughly 0.01 of finding a ULX above $10^{39}$~\ergs\ within 0.12\arcsec\ at the center of NGC 247. However, such estimate may be highly uncertain because all objects in their sample are non-nuclear sources by definition. Also, most ULXs were found to have a displacement with star clusters \citep{Kaaret2004}, suggesting a lower probability of finding a ULX associated with a star cluster.


An alternative explanation is that  the X-ray flare is due to accretion onto a nuclear black hole of the galaxy, by a tidally disrupted star or other source of gas. In the scenario of a TDE, part of the disrupted stellar materials will be accreted onto the black hole and produce electromagnetic radiation. Normal TDEs (not jetted ones) may have two characteristic features in their early phase, that they may follow a power-law decay in luminosity with an index of  $-5/3$ \citep[$L \propto t^{-5/3}$;][]{Rees1988,Komossa1999,Halpern2004}, and their emission is peaked in the UV and/or soft X-ray band \citep[cf.][]{Komossa2012} as predicted by the standard accretion disk model. This is in general consistent with observations from the current, limited sample, in which most, if not all, TDEs were recorded in their early phase and the search strategy favored objects bright in UV and soft X-rays. However, the non-detection in the UV band, with a UV luminosity at least one order of magnitude lower than that of the star cluster, i.e., $<$ $\sim$$10^{38}$~\ergs, stands out as an argument against the early-phase TDE explanation. 

The source displayed an overall decreasing trend in flux (Figure~\ref{fig:lc}). However, the short baseline of the monitoring program does not allow for a test for the $t^{-5/3}$ power-law decay model, as short-term variability is always associated with accretion system. Some TDE candidates did show strong short timescale variability superposed on a $t^{-5/3}$ long-term temporal evolution \citep[e.g., SDSS J1201+30;][]{Saxton2012}.  The X-ray energy spectrum is hard, unlike canonical TDE candidates. This could be possible if we had detected it in a late phase. Due to the absence of sensitive observations between 2011 February and 2014 July, we may have caught it 1--3 years after the initial burst. In this case, the current accretion rate is rather low, the disk is no longer a standard disk, and the emergent spectrum may have become hard.  In fact, the TDE candidate IGR J12580+0134 associated with NGC 4845 showed a hard spectrum at its luminosity peak \citep{Nikoajuk2013}. Also, the jetted TDE candidates manifested themselves as an extreme subclass with complex spectral and temporal behaviors \citep{Levan2011,Cenko2012}. These indicate that TDEs may be a diverse population.

The upper limit obtained from the {\it MAXI} data indicates that the black hole mass assuming an Eddington limited TDE is less than $\sim$$10^3$ $M_\sun$, if it had emitted a hard spectrum ($\Gamma \lesssim 2$) from the very beginning. If the initial X-ray spectrum is much softer and peaked at energies below 2 keV, then the upper limit in luminosity is much higher, e.g.,  $\sim 10^{43}$ \ergs\ given a cool disk emission with $kT_{\rm in} \approx 0.1$ keV, and so is the allowed black hole mass (up to $\sim$$10^5$ $M_\sun$). Therefore, the Eddington-limited-TDE interpretation may be possible only if the source had transitioned from a luminous soft state to a low hard state, and the central compact object is a low-mass black hole less than a few times $10^5$~$M_\sun$. Again, TDEs may be significantly sub-Eddington at their luminosity peaks, down to $\sim 10^{-2}$ even assuming the lowest black hole mass estimation \citep{Li2002,Esquej2008}. In that case, the upper limit on the black hole mass estimated with {\it MAXI} data could increase by two orders of magnitude. 

The two known TDE candidates associated with dwarf galaxies, WINGS~J1348 \citep{Maksym2013,Donato2014} and RBS~1032 \citep{Maksym2014a}, were both detected at a peak luminosity close to $10^{43}$~\ergs\ and had undergone a dramatic flux decay by a factor of about two orders of magnitude. WINGS~J1348 displayed a luminosity around $10^{41}$~\ergs\ 6 years after its discovery, and so did RBS~1032 almost 20 years after its first detection. They emitted a soft blackbody-like spectrum throughout the active cycle with a temperature near 100 eV. The current activity of the object in NGC 247 cannot be longer than 3.5 years, but its spectrum has been hard and the luminosity has gone below $10^{39}$~\ergs. Thus, this object is at least not in the same class as the two TDEs found in dwarfs.

X-ray observations of the Galactic center molecular clouds indicate that the central supermassive black hole Sgr~A$^\star$ may also have exhibited giant X-ray flares of similar luminosities in the past a few centuries \citep[for a review see][]{Ponti2013}. Specifically, the X-ray emission from the molecular cloud Sgr~B2 requires X-ray flaring activities from Sgr~A$^\star$ which have a power-law spectrum with a photon index $\Gamma \approx 2$ and an X-ray luminosity of roughly $10^{39}$ \ergs\ lasting a few years \citep[e.g.][]{Terrier2010}. The spectrum and luminosity are similar to those observed for NGC 247. In addition to the possibility of being triggered by a tidally disrupted star, the accretion materials could come from a stream of gas similar to the case of the G2 cloud near Sgr~A$^\star$ \citep{Pfuhl2015}, in which case the accretion rate may be insufficient to reach the Eddington limit.


Thus, the temporal and spectral characteristics of this object can be well explained by a stellar-mass black hole accreting close to the Eddington limit at its luminosity peak. However, we emphasize that the possibility for nuclear black hole activity cannot be completely ruled out. The absence of sensitive X-ray observations in the early phase of the flare hampers a conclusive diagnosis of the source nature. Also, the TDE behavior may be too complex to predict especially in its late phase. Plus, the trigger could be other sources like a small gas cloud rather than a disrupted star, whose consequent accretion behavior is even more difficult to predict. A continued X-ray monitoring will be a key to distinguishing the different scenarios. If this object is proven to be a nuclear black hole, it will have great interest in two aspects. First, its proximity will offer us a unique test laboratory for TDE physics at the late phase, if it is a TDE. Also, it suggests that monitoring in X-rays for flaring activity may be an effective tool for us to unveil hidden nuclear black holes in nearby quiescent galaxies. 

\acknowledgements 
We thank the anonymous referee for helpful comments that improved the paper, the {\it Chandra} director for granting discretionary time and the {\it Swift} PI for approving the ToO program, and the {\it XMM-Newton}/{\it Chandra}/{\it Swift} teams for successful executions of the observations. HF acknowledges funding support from the National Natural Science Foundation of China under grant No.\ 11222327, and the Tsinghua University Initiative Scientific Research Program. LCH acknowledges support by the Chinese Academy of Science through grant No.\ XDB09030102 (Emergence of Cosmological Structures) from the Strategic Priority Research Program and by the National Natural Science Foundation of China through grant No.\ 11473002.

{\it Facilities:} \facility{XMM-Newton}, \facility{Chandra}, \facility{Swift}


\begin{thebibliography}{}

\bibitem[Bachetti et al.(2014)]{Bachetti2014}
Bachetti, M., Harrison, F.~A., Walton, D.~J., et al.\ 2014, \nat, 514, 202 

\bibitem[Barth et al.(2008)]{Barth2008}
Barth, A.~J., Greene, J.~E., \& Ho, L.~C.\ 2008, \aj, 136, 1179 

\bibitem[Barth et al.(2004)]{Barth2004}
Barth, A.~J., Ho, L.~C., Rutledge, R.~E., \& Sargent, W.~L.~W.\ 2004, \apj, 607, 90 

\bibitem[Brunthaler et al.(2009)]{Brunthaler2009}
Brunthaler, A., Castangia, P., Tarchi, A., et al.\ 2009, \aap, 497, 103 

\bibitem[Buccheri et al.(1983)]{Buccheri1983}
Buccheri, R., Bennett, K., Bignami, G.~F., et al.\ 1983, \aap, 128, 245 

\bibitem[Carignan(1985)]{Carignan1985}
Carignan, C.\ 1985, \apjs, 58, 107 

\bibitem[Carignan \& Puche(1990)]{Carignan1990}
Carignan, C., \& Puche, D.\ 1990, \aj, 100, 641 

\bibitem[Cenko et al.(2012)]{Cenko2012}
Cenko, S.~B., Krimm, H.~A., Horesh, A., et al.\ 2012, \apj, 753, 77 

\bibitem[Davidge(2010)]{Davidge2010}
Davidge, T.~J.\ 2010, \apj, 725, 1342 

\bibitem[Davidge \& Courteau(2002)]{Davidge2002}
Davidge, T.~J., \& Courteau, S.\ 2002, \aj, 123, 1438 

\bibitem[Desroches \& Ho(2009)]{Desroches2009}
Desroches, L.-B., \& Ho, L.~C.\ 2009, \apj, 690, 267 

\bibitem[de Vaucouleurs et al.(1991)]{deVaucouleurs1991}
de Vaucouleurs, G., de Vaucouleurs, A., Corwin, H.~G., Jr., et al.\ 1991, Third
Reference Catalogue of Bright Galaxies,~Vol. 3 (New York: Springer), 467

\bibitem[Donato et al.(2014)]{Donato2014}
Donato, D., Cenko, S.~B., Covino, S., et al.\ 2014, \apj, 781, 59 

\bibitem[Dong et al.(2012)]{Dong2012}
Dong, X.-B., Ho, L.~C., Yuan, W., et al.\ 2012, \apj, 755, 167 

\bibitem[Donley et al.(2002)]{Donley2002}
Donley, J.~L., Brandt, W.~N., Eracleous, M., \& Boller, T.\ 2002, \aj, 124, 1308 

\bibitem[Esquej et al.(2007)]{Esquej2007}
Esquej, P., Saxton, R.~D., Freyberg, M.~J., et al.\ 2007, \aap, 462, L49 

\bibitem[Esquej et al.(2008)]{Esquej2008}
Esquej, P., Saxton, R.~D., Komossa, S., et al.\ 2008, \aap, 489, 543 

\bibitem[Ferguson et al.(1996)]{Ferguson1996}
Ferguson, A.~M.~N., Wyse, R.~F.~G., Gallagher, J.~S., III, \& Hunter, D.~A.\ 1996, \aj, 111, 2265 

\bibitem[Filippenko \& Ho(2003)]{Filippenko2003}
Filippenko, A.~V., \& Ho, L.~C.\ 2003, \apjl, 588, L13 

\bibitem[Gallo et al.(2010)]{Gallo2010}
Gallo, E., Treu, T., Marshall, P.~J., et al.\ 2010, \apj, 714, 25 

\bibitem[Gladstone et al.(2009)]{Gladstone2009}
Gladstone, J.~C., Roberts, T.~P., \& Done, C.\ 2009, \mnras, 397, 1836 

\bibitem[Gieren et al.(2009)]{Gieren2009}
Gieren, W., Pietrzy{\'n}ski, G., Soszy{\'n}ski, I., et al.\ 2009, \apj, 700, 1141 

\bibitem[Greene(2012)]{Greene2012}
Greene, J.~E.\ 2012, Nature Communications, 3, 1304 

\bibitem[Greene \& Ho(2004)]{Greene2004}
Greene, J.~E., \& Ho, L.~C.\ 2004, \apj, 610, 722 

\bibitem[Greene \& Ho(2007a)]{Greene2007a} 
Greene, J.~E., \& Ho, L.~C.\ 2007a, \apj, 667, 131 

\bibitem[Greene \& Ho(2007b)]{Greene2007b} 
Greene, J.~E., \& Ho, L.~C.\ 2007b, \apj, 670, 92 

\bibitem[Halpern et al.(2004)]{Halpern2004}
Halpern, J.~P., Gezari, S., \& Komossa, S.\ 2004, \apj, 604, 572 

\bibitem[Ho(2008)]{Ho2008}
Ho, L.~C.\ 2008, \araa, 46, 475 

\bibitem[Izotov \& Thuan(2008)]{Izotov2008}
Izotov, Y.~I., \& Thuan, T.~X.\ 2008, \apj, 687, 133 

\bibitem[Izotov et al.(2012)]{Izotov2012}
Izotov, Y.~I., Thuan, T.~X., \& Privon, G.\ 2012, \mnras, 427, 1229 

\bibitem[Jin et al.(2011)]{Jin2011}
Jin, J., Feng, H., Kaaret, P., \& Zhang, S.-N.\ 2011, \apj, 737, 87 

\bibitem[Kaaret et al.(2004)]{Kaaret2004}
Kaaret, P., Alonso-Herrero, A., Gallagher, J.~S., et al.\ 2004, \mnras, 348, L28 

\bibitem[Kalberla et al.(2005)]{Kalberla2005}
Kalberla, P.~M.~W., Burton, W.~B., Hartmann, D., et al.\ 2005, \aap, 440, 775 

\bibitem[Komossa(2012)]{Komossa2012}
Komossa, S.\ 2012, European Physical Journal Web of Conferences, 39, 02001 

\bibitem[Komossa \& Bade(1999)]{Komossa1999}
Komossa, S., \& Bade, N.\ 1999, \aap, 343, 775 

\bibitem[Kormendy \& Ho(2013)]{Kormendy2013}
Kormendy, J., \& Ho, L.~C.\ 2013, \araa, 51, 511 

\bibitem[La Parola et al.(2003)]{LaParola2003}
La Parola, V., Damiani, F., Fabbiano, G., \& Peres, G.\ 2003, \apj, 583, 758 

\bibitem[Lasota(2001)]{Lasota2001}
Lasota, J.-P.\ 2001, NewAR, 45, 449 

\bibitem[Leahy et al.(1983)]{Leahy1983}
Leahy, D.~A., Darbro, W., Elsner, R.~F., et al.\ 1983, \apj, 266, 160 

\bibitem[Lemons et al.(2015)]{Lemons2015}
Lemons, S., Reines, A., Plotkin, R., Gallo, E., \& Greene, J.\ 2015, \apj, in press

\bibitem[Levan et al.(2011)]{Levan2011} 
Levan, A.~J., Tanvir, N.~R., Cenko, S.~B., et al.\ 2011, Science, 333, 199 

\bibitem[Li et al.(2002)]{Li2002}
Li, L.-X., Narayan, R., \& Menou, K.\ 2002, \apj, 576, 753 

\bibitem[Luo et al.(2008)]{Luo2008}
Luo, B., Brandt, W.~N., Steffen, A.~T., \& Bauer, F.~E.\ 2008, \apj, 674, 122 

\bibitem[Maksym et al.(2013)]{Maksym2013}
Maksym, W.~P., Ulmer, M.~P., Eracleous, M.~C., Guennou, L., \& Ho, L.~C.\ 2013, \mnras, 35, 1904 

\bibitem[Maksym et al.(2014a)]{Maksym2014a} 
Maksym, W.~P., Lin, D., \& Irwin, J.~A.\ 2014a, \apjl, 792, L29 

\bibitem[Maksym et al.(2014b)]{Maksym2014b} 
Maksym, W.~P., Ulmer, M.~P., Roth, K.~C., et al.\ 2014b, \mnras, 444, 866 

\bibitem[Maoz et al.(1995)]{Maoz1995}
Maoz, D., Filippenko, A.~V., Ho, L.~C., et al.\ 1995, \apj, 440, 91 

\bibitem[Matsuoka et al.(2009)]{Matsuoka2009}
Matsuoka, M., Kawasaki, K., Ueno, S., et al.\ 2009, PASJ, 61, 999 

\bibitem[Mineshige et al.(1994)]{Mineshige1994}
Mineshige, S., Hirano, A., Kitamoto, S., Yamada, T.\ T., \& Fukue, J.\ 1994, \apj, 426, 308

\bibitem[Motch et al.(2014)]{Motch2014}
Motch, C., Pakull, M.~W., Soria, R., Gris{\'e}, F., \& Pietrzy{\'n}ski, G.\ 2014, \nat, 514, 198 

\bibitem[Niko{\l}ajuk \& Walter(2013)]{Nikoajuk2013}
Niko{\l}ajuk, M., \& Walter, R.\ 2013, \aap, 552, A75 

\bibitem[Ott et al.(2012)]{Ott2012}
Ott, J., Stilp, A.~M., Warren, S.~R., et al.\ 2012, \aj, 144, 123 

\bibitem[Peterson et al.(2005)]{Peterson2005}
Peterson, B.~M., Bentz, M.~C., Desroches, L.-B., et al.\ 2005, \apj, 632, 799 

\bibitem[Pfuhl et al.(2015)]{Pfuhl2015}
Pfuhl, O., Gillessen, S., Eisenhauer, F., et al.\ 2015, \apj, 798, 111 

\bibitem[Pierce \& Tully(1992)]{Pierce1992}
Pierce, M.~J., \& Tully, R.~B.\ 1992, \apj, 387, 47 

\bibitem[Ponti et al.(2013)]{Ponti2013}
{Ponti}, G., {Morris}, M.~R., {Terrier}, R., \& {Goldwurm}, A. 2013, in Advances in Solid State Physics, Vol.~34, Cosmic Rays in Star-Forming Environments, ed. D.~F. {Torres} \& O.~{Reimer}, 331 

\bibitem[Rees(1988)]{Rees1988}
Rees, M.~J.\ 1988, \nat, 333, 523 

\bibitem[Reines et al.(2011)]{Reines2011}
Reines, A.~E., Sivakoff, G.~R., Johnson, K.~E., \& Brogan, C.~L.\ 2011, \nat, 470, 66 

\bibitem[Remillard \& McClintock(2006)]{Remillard2006}
Remillard, R.~A., \& McClintock, J.~E.\ 2006, \araa, 44, 49 

\bibitem[Saxton et al.(2012)]{Saxton2012}
Saxton, R.~D., Read, A.~M., Esquej, P., et al.\ 2012, \aap, 541, A106 

\bibitem[Schlegel et al.(1998)]{Schlegel1998}
Schlegel, D.~J., Finkbeiner, D.~P., \& Davis, M.\ 1998, \apj, 500, 525 

\bibitem[Seth et al.(2010)]{Seth2010}
Seth, A.~C., Cappellari, M., Neumayer, N., et al.\ 2010, \apj, 714, 713 

\bibitem[Shakura \& Sunyaev(1973)]{Shakura1973}
Shakura, N.\ I., \& Sunyaev, R.\ A.\ 1973, \aap, 24, 337

\bibitem[Skinner et al.(1982)]{Skinner1982}
Skinner, G.~K., Bedford, D.~K., Elsner, R.~F., et al.\ 1982, \nat, 297, 568 

\bibitem[Stobbart et al.(2006)]{Stobbart2006}
Stobbart, A.-M., Roberts, T.~P., \& Wilms, J.\ 2006, \mnras, 368, 397 

\bibitem[Sutton et al.(2013)]{Sutton2013}
Sutton, A.~D., Roberts, T.~P., \& Middleton, M.~J.\ 2013, \mnras, 435, 1758 

\bibitem[Swartz et al.(2011)]{Swartz2011}
Swartz, D.~A., Soria, R., Tennant, A.~F., \& Yukita, M.\ 2011, \apj, 741, 49 

\bibitem[Tao et al.(2012)]{Tao2012}
Tao, L., Feng, H., Kaaret, P., Gris{\'e}, F., \& Jin, J.\ 2012, \apj, 758, 85 

\bibitem[{Terrier} {et~al.}(2010)]{Terrier2010}
{Terrier}, R., {Ponti}, G., {B{\'e}langer}, G., {et~al.} 2010, \apj, 719, 143

\bibitem[Wang \& Merritt(2004)]{Wang2004}
Wang, J., \& Merritt, D.\ 2004, \apj, 600, 149 

\bibitem[Wilms et al.(2000)]{Wilms2000}
Wilms, J., Allen, A., \& McCray, R.\ 2000, \apj, 542, 914 

\end{thebibliography}
\end{document}